\RequirePackage[2020-02-02]{latexrelease}
\documentclass[aps,floatfix]{revtex4}
\usepackage{amsmath}
\usepackage{centernot}
\usepackage{graphics,epsfig}
\usepackage{graphicx}
\begin{document}
\title{Israel coordinates for all static spherically symmetric spacetimes \linebreak
with vanishing second Ricci invariant}
\author{Yannick M. Bisson \cite{emailyb} and Kayll Lake \cite{emailkl}}
\affiliation{Department of Physics, Queen's University, Kingston, Ontario, Canada, K7L3N6 }
\date{\today}
\begin{abstract}
Static spherically symmetric spacetimes with vanishing second Ricci invariant constitute an important class of solutions to Einstein's equations and more generally as archetypes of regular black holes.  When studying completeness one is most often presented with the Kruskal - Szekeres procedure. However, this procedure only works if the spacetime admits a single non-degenerate Killing horizon (a single bifurcation two-sphere). Here we generalize the Israel procedure to examine a constructive approach to completeness based entirely on the static spherically symmetric nature of spacetimes with a vanishing second Ricci invariant. It is shown by ``block gluing" that the Israel procedure can cover two bifurcation two-spheres, but can fail with three. No coordinate transformations are used in this work.
\end{abstract}
\maketitle
\section{Introduction}

The metrics \cite{conventions}
\begin{equation}
ds^2=-f(r)dt^2+\frac{dr^2}{f(r)}+r^2d\Omega_{2}^{2},
\label{fstatic}
\end{equation}
where $d\Omega_{2}^{2}$ is the metric of a unit $2$-sphere, constitute a very well-known class of solutions to Einstein's equations, and, depending on the form of $f(r)$, allow simple models for regular black holes. Some properties of the metrics (\ref{fstatic}) have been studied by Jacobson \cite{Jacobson}. More recently, the metrics (\ref{fstatic}) have been invariantly characterized by the vanishing of their second Ricci invariant ($R_{2}$ defined below) \cite{Lake}. Moreover, as is well known, the metrics (\ref{fstatic}) posses the hypersurface - orthogonal Killing vectors $\xi^{\alpha} = \delta^{\alpha}_{t}$ where $\xi{^\alpha}\xi_{\alpha} = -f(r)$. We use these invariant properties in the development which follows \cite{footnote1}. Usually one is interested in the complete manifold associated with (\ref{fstatic}). In the case of the Penrose - Carter procedure the solution to this problem via ``block gluing" in a conformally related space has been available for many years \cite{Walker}. More general block gluing constructions are given in \cite{piotr}. However, when one turns to complete coordinate representations of (\ref{fstatic}), the situation is quite different. Usually, one is introduced to the Kruskal \cite{Kruskal} - Szekeres \cite{Szekeres} procedure \cite{Lake1}. However, this procedure only works for a single simple root: there exists a single $r_{0}$ such $f(r_{0}) = 0$ with $f'(r_{0}) \neq 0$. The purpose of this communication is to offer a different construction which works in a wider class of situations. We show that the Israel procedure covers more cases than the Kruskal - Szekeres procedure, but there are cases when the Israel coordinates remain incomplete. This incompleteness is shown by way of the block gluing procedure. No coordinate transformations are used in this work, nor are any field equations.
\section{Generalized Israel Coordinates}
\subsection{General Properties}
We start with a spherically symmetric spacetime in coordinates $(u,w,\theta,\phi)$ where
$k^{\alpha}=\delta^{\alpha}_{w}$ is a radial null vector so that the line
element takes the form \cite{Lake2}
\begin{equation}
ds^2=\mathcal{F}(u,w)du^2+2h(u,w)dudw+r(u,w)^2d\Omega_{2}^{2}.
\label{generalmetric}
\end{equation}
Further, setting $k^{\beta}\nabla_{\beta}k^{\alpha}=0$ (so that
trajectories with tangents $k$ are radial null geodesics affinely
parameterized by $w$) it follows that $\partial h/\partial w=0$.
We retain $h(u)$ in this section for possible future convenience.
Note that the range in $u$ is $-\infty <u< \infty$ and over this range it is assumed that the
associated null geodesics cover all of the spacetime.

The expansion of $k^{\alpha}$ is given by
\begin{equation}
\nabla_{\alpha}k^{\alpha}= \frac{2}{r}r_{w}, \label{expansiongendg4}
\end{equation}
where a coordinate subscript now represents partial differentiation.

Consider the 4-vector
\begin{equation}
l^{\alpha}\partial_{\alpha} = 2h\partial_{u}-\mathcal{F}\partial_{w}. \label{lvector}
\end{equation}
We find that $l^{\alpha}l_{\alpha} = 0$ and that $l^{\beta}\nabla_{\beta}l^{\alpha} = \kappa l^{\alpha}$ where
\begin{equation}
\kappa = 4 h'-\mathcal{F}_{w}, \label{kappa}
\end{equation}
and $' \equiv d/du$ so that $l^{\alpha}$ is tangent to a non-affinely parameterized radial null geodesic. The apparent horizon is distinguished by the condition $\nabla_{\alpha}l^{\alpha} = \kappa$ \cite{poisson} which requires
\begin{equation}
2h r_{u} = \mathcal{F}r_{w}. \label{apparent}
\end{equation}
\subsection{The Second Ricci Invariant}
Up to a physically irrelevant numerical coefficient the second Ricci invariant is given by \cite{Lake}
\begin{equation}
R_{2} \equiv S^{\alpha}_{\beta}S^{\gamma}_{\alpha}S^{\beta}_{\gamma} \label{R2}
\end{equation}
where the trace - free Ricci tensor $S^{\alpha}_{\beta}$ is given by
\begin{equation}
S^{\alpha}_{\beta} =  R^{\alpha}_{\beta} -\frac{R}{4} \delta^{\alpha}_{\beta}\label{S}
\end{equation}
where $R^{\alpha}_{\beta}$ is the Ricci tensor, $R$ the Ricci scalar and $\delta^{\alpha}_{\beta}$ the Kronecker delta.
It is adequate for our purposes here to set $h$ to a constant. (A preferred value of this constant is given in the next section.) Then, with the aid of \textit{GRTensorIII} \cite{grt} we find
\begin{equation}
R_{2} \propto \frac{R_{2a}R_{2b}R_{2c}}{r^4}\label{R2form}
\end{equation}
where
\begin{equation}
R_{2a} \equiv r_{ww}, \label{R2a}
\end{equation}
\begin{equation}
R_{2b} \equiv \mathcal{F}_{ww}r^2-2\mathcal{F}r_{w}^2+4r_{u}r_{w}h-2h^2, \label{R2b}
\end{equation}
and
\begin{equation}
R_{2c} \equiv r_{ww}\mathcal{F}^2-4hr_{uw}\mathcal{F}-2\mathcal{F}_{u}r_{w}h+2\mathcal{F}_{w}r_{u}h+4r_{uu}h^2. \label{R2c}
\end{equation}
Clearly
\begin{equation}
R_{2a}=0 \Rightarrow r(u,w)=f_{1}(u)w+f_{2}(u).\label{R2aa}
\end{equation}
It is easy to obtain misinformation on the relations $R_{2b} = 0$ and $R_{2c} = 0$ \cite{maplesoft}.
However, to proceed, it is essential that we first seek Killing vectors since non-static cases are known with $R2 = 0$ \cite{Lake}. As explained in the next section, we conclude, without loss in generality, that \cite{f}
\begin{equation}
 r(u,w)=f_{1}w+f_{2},\label{R2zero}
\end{equation}
 $f_{1} \neq 0$ \cite{f10}.
\subsection{Killing Vectors}
We now seek hypersurface - orthogonal Killing vectors. Specifically, we seek radial 4-vectors $\xi_{\mu}$ such that
\begin{equation}
 \nabla_{\mu}\xi_{\nu} +  \nabla_{\nu}\xi_{\mu} \equiv \Xi_{\mu \nu} = 0 \label{killing1}
\end{equation}
and
\begin{equation}
 \xi_{[\alpha}\nabla_{\mu}\xi_{\nu]} = 0.\label{killing2}
\end{equation}
We do not impose equation (\ref{R2zero}) apriori but retain the $f_1$ $f_2$ notation for convenience.

Writing
\begin{equation}
\xi^{\alpha}\partial_{\alpha} = A(u,w)\partial_{u}+B(u,w)\partial_{w}\label{AB}
\end{equation}
it follows that (\ref{killing2}) is satisfied for all smooth $A$ and $B$.
Next, setting $\Xi_{w w}=0$ we find that
\begin{equation}
A(u,w) = f_1(u), \label{ww}
\end{equation}
and with (\ref{ww}) that $\Xi_{uw} = 0$ for
\begin{equation}
B(u,w) = -f_1'(u)w-f_2'(u) \label{uw}
\end{equation}
and with (\ref{uw}) that $\Xi_{\theta \theta} = 0$ for
\begin{equation}
r(u,w) = J(f_1w+f_2) \label{theta}
\end{equation}
were $J$ is any suitably smooth function, not necessarily the identity function. Combining (\ref{ww}) and (\ref{uw}) we find that $\Xi_{uu} = 0$ for
\begin{equation}
\mathcal{F} = 2h \frac{f_1'w + f_2'}{f_{1}} - \frac{F(f_1w+f_2)}{f_{1}^2}\label{F(u,w)}
\end{equation}
where $F$ is any suitably smooth function, in general distinct from $J$. Whereas the solution (\ref{R2zero}) corresponds to the identity function for $J$, this does not change (\ref{F(u,w)}). 

\bigskip

With the aide of (\ref{theta}) and (\ref{F(u,w)}) it follows from (\ref{R2b}) with $R_{2b} = 0$ that the two functions $J$ and $F$ are related by the differential relation
\begin{equation}
-\frac{d^2F}{dx^2}J^2+ 2(\frac{dJ}{dx})^2F = 2h^2. \label{fjb}
\end{equation}
To proceed, $F$ or $J$ or a relationship between them must be given \cite{identity}. No such information is available. Further, with the aide of (\ref{theta}) and (\ref{F(u,w)}) it follows from (\ref{R2c}) with $R_{2c} = 0$ that the two functions $J$ and $F$ are related by the differential relation
\begin{equation}
\frac{d^2J}{dx^2}F^2 = 0. \label{fjc}
\end{equation}
In this case if $J$ is chosen as the identity function (\ref{fjc}) gives $0 = 0$ for all $F$. If $J$ is not the identity function then $F = 0$ which is clearly unacceptable. We conclude that (\ref{R2a}) and (\ref{R2c}) give (\ref{R2zero}) but no useful information comes from (\ref{R2b}). At this point $F(r)$ is an arbitrary but smooth function.

Further information about $F$ can be obtained by considering the (invariant) Hernandez - Misner mass \cite{hm}
\begin{equation}
\mathcal{M} \equiv \frac{r}{2}R_{\theta \phi}^{\;\;\;\; \theta \phi}\label{mhm}
\end{equation}
where $R$ is the Riemann tensor. From (\ref{fstatic}) and (\ref{generalmetric}) with (\ref{F(u,w)}) and the invariance we find that \cite{M}
\begin{equation}
F = f \label{Ff}
\end{equation}
for $h^2 = 1$, a convenience which sets our choice for $h^2$.

In summary, the 4-vector
\begin{equation}
\xi^{\alpha}\partial_{\alpha} = C(-r_{w}\partial_{u}+r_{u}\partial_{w}), \label{FK}
\end{equation}
given (\ref{R2zero}) and (\ref{F(u,w)}), satisfies (\ref{killing1}) and (\ref{killing2}). It follows that
\begin{equation}
\xi^{\alpha}\xi_{\alpha} = -C^2F, \label{FKM}
\end{equation}
a well-known fact in (\ref{fstatic}) (given (\ref{Ff})) now transposed to (\ref{generalmetric}) without coordinate transformation. Of course ``static" refers to timelike $\xi^{\alpha}$, regions for which $F>0$.
\section{Familiar Examples}
\subsection{Generalized Eddington - Finkelstein Coordinates}
Take
\begin{equation}
f_{1} = 1, f_{2} = 0. \label{EFff}
\end{equation}
Then, (\ref{generalmetric}) takes the form
\begin{equation}
ds^2=-(1-\frac{2 \mathcal{M}(r)}{r})du^2\pm2dudr+r^2d\Omega_{2}^{2}
\label{EF}
\end{equation}
with $r= w$. This is the generalized Eddington - Finkelstein form. The coordinates are well-known to be incomplete.
\subsection{Original Israel Coordinates}
Take
\begin{equation}
f_{1} = \frac{hu}{4m}, f_{2} = 2m, \label{If}
\end{equation}
where $m$ is a constant, so that $r = \frac{huw}{4m} + 2m$.
Then, (\ref{generalmetric}) takes the form
\begin{equation}
ds^2=(\frac{w^2}{2mr})du^2+2hdudw+r^2d\Omega_{2}^{2}.
\label{israel}
\end{equation}
This is the Israel form of the Schwarzschild metric \cite{Israel} (he chose $h = +1$). The coordinates are known to be complete. See also \cite{Newman} and \cite{Klosch}. Note that in the context of this work there is no relation between the $u$ used in (\ref{EF}) and the $u$ used in (\ref{israel}) as, once again, no coordinate transformations have been used.
\section{More General Situations}
We now turn to invariants. For the spacetimes under consideration here, given the requirement $R2 = 0$, it is known that there remain only three independent scalar invariants derivable from the Riemann tensor without differentiation. These are the Ricciscalar $R$, the first Ricci invariant $R1$, and the first Weyl invariant $W1R$ (see the reference in \cite{footnote1}). For \textit{all} choices of $f_{1}$ and $f_{2}$, where now $' \equiv d/dr$, these are given, up to irrelevant numerical factors, by

\begin{equation}
W1R \propto \frac{1}{r^4}(F''r^2-2F'r+2F-2)^2,
\label{Weyl1}
\end{equation}

\begin{equation}
R = \frac{1}{r^2}(-F''r^2-4F'r-2F+2),
\label{Ricci}
\end{equation}

and

\begin{equation}
R1 \propto \frac{1}{r^4}(F''r^2-2F+2)^2.
\label{Ricci1}
\end{equation}

\bigskip

There are two obvious ways to proceed: \textit{(i)} We can impose conditions on the invariants and solve for $F$, or \textit{(ii)} We can impose restrictions on $F$ which, for example, render the invariants regular. As an example of the first case, setting $R = 4 \Lambda$, where $\Lambda$ is a constant, the resultant differential equation can be solved to give
\begin{equation}
F = 1 + \frac{c_{1}}{r} + \frac{c_{2}}{r^2} - \frac{\Lambda r^2}{3},
\label{rnds}
\end{equation}
where $c_{1}$ and $c_{2}$ are constants. In Einstein's theory these are the Reissner - Nordstr\"{o}m - de Sitter solutions (for $\Lambda > 0$) though we have no reason to associate $c_{2}$ with charge here. The cases $c_{2} = 0$ have been studied in detail previously \cite{Lake2}. For the case $c_{2} \neq 0$ (but $\Lambda =0$) see \cite{Klosch}. Unlike the Kruskal - Szekeres procedure, the generalized Israel coordinates can handle two distinct roots to $F = 0$ \cite{ksfailure}. However, (\ref{rnds}) shows that the Israel coordinates can fail. If none of $c_{1}, c_{2}$ and $\Lambda$ are zero, then there can be three distinct Killing horizons and the associated conformal block diagram (see \cite{piotr}) shows that the coordinate $u$, even over the range $-\infty <u< \infty$, fails to access the entire spacetime.

\bigskip

As regards regularity of the invariants, we first observe that for $r \geq 0$ and $F \in C^2$ the invariants can possibly diverge only at $r = 0$. From the forms given it follows immediately that the spacetimes are regular for $F(0) = 1$ and $F'(0) = 0$. These are a special case of regularity conditions known for many years \cite{lakemus}. Regularity of the invariants brings us to the somewhat murky area of "regular" black holes. We say murky because more often than not the invariants to be considered are either not known or not explained though the problem was completely solved in the spherically symmetric case in \cite{footnote1}. Further, one sometimes sees statements like ``curvature invariants do not have a real physical meaning" (\textit{e.g.} \cite{modesto}). As explain in \cite{Lake}, this is incorrect. Simply use Einstein's equations in the Ricci invariants.
\begin{acknowledgments}
This work was supported by a grant from the Natural Sciences and
Engineering Research Council of Canada and was made possible by
use of \textit{GRTensorIII} \cite{grt}. 
\end{acknowledgments}


\begin{thebibliography}{}
\bibitem{conventions}We use geometrical units and a signature of $+2$. Functional dependence is usually deignated only on the first appearance of a function. Nowadays, with the wide availability of computer algebra programs, which are a prerequisite to our calculations, we feel no obligation to record the long intermediate expressions that arise in parts of this work.
\bibitem{footnote1}It is to be noted that the static form (\ref{fstatic}) is not unique in the sense that the form
\begin{equation*}
ds^2=-a(r)dt^2+b(r)dr^2+r^2d\Omega_{2}^{2}.
\end{equation*}
is also static.
Note that if $b(r) \neq 1/a(r)$ and $\exists \; r = r_{0} \ni a(r_{0}) = 0$ then it follows that there is a singularity at $r_{0}$ as all non-differential invariants $ \rightarrow \infty$ unless perhaps if $a'(r_{0}) = 0$. We do not consider these cases here. See K. Santosuosso, D. Pollney, N. Pelavas, P. Musgrave and K. Lake, Computer Physics Communications \textbf{115}, 381 (1998) (gr-qc/9809012).
\bibitem{f} All functions $f_n(u)$ are considered suitably smooth.
\bibitem{f10} Specifically, we mean $f_{1}(u) \neq 0$ not ruling out isolated zeros $f_{1}(u_{0}) = 0$. An example of what we do not consider here is the Bertotti - Kasner metric (see \cite{Lake2}).
\bibitem{identity} For example, if $J$ is chosen as the identity function, it follows that $F(x)=h^2+c1/x+c2x^2$ for constants $c1$ and $c2$. This is clearly too restrictive.
\bibitem{ksfailure} The reason for these distinct results is easy to see. In the Kruskal - Szekeres procedure $r = r(uv)$ where trajectories of constant $u$ or $v$ are radial null geodesics. There can be but one bifurcation two - sphere $r=r(0)$ ($r$ is sigle valued at $uv = 0$). In the Israel coordinates $r = r(u,w)$ where trajectories of constant $u$ are radial null geodesics affinely parameterized by $w$. By inspection of the Penrose - Carter diagram it is easy to see that two bifurcation two - spheres can be covered. See also \cite{Lake2} for a distinct case with two bifurcation two - spheres.
\end{thebibliography}
\end{document}